\documentclass[doublecol,graphics]{epl2} 
\usepackage{amsmath}
\usepackage{amssymb}
\usepackage{color}

\newcommand{\be}{\begin{equation}}
\newcommand{\ee}{\end{equation}}
\newcommand{\bd}{\begin{displaystyle}}
\newcommand{\ed}{\end{displaystyle}}
\newcommand{\ba}{\begin{eqnarray}}
\newcommand{\ea}{\end{eqnarray}}
\newcommand{\ket}[1]{{| #1 \rangle}}

\newcommand{\aver}[1]{{\langle #1 \rangle}}
\DeclareMathOperator{\Rre}{Re}
\DeclareMathOperator{\Iim}{Im}

\title{Nonlinear control of chaotic walking of atoms in an optical lattice}
\shorttitle{Nonlinear control of chaotic walking of atoms in an optical lattice}

\author{V.\,Yu. Argonov \and S.\,V. Prants}
\shortauthor{V.\,Yu. Argonov and S.\,V. Prants}

\institute{   
Laboratory of Nonlinear Dynamical Systems,\\Pacific Oceanological
Institute of the Russian Academy of Sciences,\\43 Baltiiskaya st., 690041
Vladivostok, Russia}

\abstract{Centre-of-mass atomic motion in an optical lattice {is 
shown to be near the resonance} a chaotic walking due to the interplay 
between coherent internal
atomic dynamics and spontaneous emission. Statistical properties
of chaotic atomic motion can be controlled by the single parameter, the
detuning between the atomic transition frequency and
the laser frequency. We derive a Fokker-Planck equation in the energetic
space to describe the atomic transport near the resonance and
demonstrate numerically how to manipulate the atomic motion varying the
detuning.}

\begin{document}

\maketitle

\section{Introduction}

Let us consider a cold two-level atom
in a laser standing wave forming a 1D optical
lattice. {Depending on the values of atomic and lattice parameters, 
different regimes of the centre-of-mass motion can be identified:} 
oscillations in wells of the optical potential,
ballistic flights with acceleration and
deceleration, velocity grouping, Brownian motion
and random walking with L\'evy flights \cite{Kaz,
BBAC, MEZ, KSW, Stenholm, LCTA, JD96, KO98, RA06, GD06}.
Theoretical study and experimental realization
of these effects have been done, mainly, in the
context of laser cooling of atoms when the 
laser is far detuned from the atomic transition
frequency and the excited state can be adiabatically
eliminated. Near the resonance, coherent
interaction of atoms with a laser field may
strongly affect the atomic motion. It has been
shown in Refs. \cite{PK01, PS01, AP03} that,
when neglecting spontaneous emission (SE), there should
exist {\it a deterministic chaotic transport} of atoms 
with exponential sensitivity to small variations
in initial conditions and/or lattice
parameters. In our recent work \cite{AP07} we
developed a semiclassical theory of this phenomenon
{resembling} a
random walking but without any random forces and noise.

In this Letter we use a Monte Carlo 
wavefunction approach to take into account
SE and demonstrate a new type
of motion in an optical
lattice, {\it a chaotic atomic walking}, with the properties both of the chaotic
atomic transport, caused by coherent atomic dynamics, and of a random process due to
SE kicks. It is a kind of random walking with specific statistical properties
that {cannot} be classified neither as deterministic chaotic
motion nor as normal diffusion nor as sub(super)diffusion and
nor as L\'evy flights.  We derive a
Fokker-Planck equation in the energetic space, study
the statistical properties of the chaotic walking
and demonstrate numerically how to manipulate
these properties varying the atom-field detuning.
Small changes in this parameter may affect {drastically} 
the atomic transport, transforming the atomic motion from a 
practically regular one to anomalous diffusion.

\section{Monte Carlo wavefunction approach}

We start with the non-Hermitian Hamiltonian
of a two-level atom interacting with a strong standing-wave 1D
laser field forming an optical lattice. In the
frame rotating with the laser frequency $\omega_f$, 
it has the form
\begin{equation}
\begin{aligned}
\hat H=& \frac{\hat P^2}{2m_a}+\frac{1}{2}\hbar(\omega_a-\omega_f)\hat\sigma_z-
\\& -\hbar \Omega\left(\hat\sigma_-+\hat\sigma_+\right)\cos{k_f\hat X}-i\hbar\frac{\Gamma}{2}\hat\sigma_+\hat\sigma_-,
\label{Jaynes-Cum}
\end{aligned}
\end{equation}
where $\hat\sigma_{\pm, z}$ are the Pauli operators for the
internal atomic degrees of freedom, $\hat X$ and $\hat P$
are the atomic position and momentum operators,
$\omega_a$, $\omega_f$ and $\Omega$ are the atomic transition,
laser and Rabi frequencies, respectively, and $\Gamma$
is the spontaneous decay rate. {Internal
atomic states are} described by the wavefunction
$\ket{\Psi(t)}=a(t)\ket{2}+b(t)\ket{1}$, with $a$ and $b$ being the
complex-valued probability amplitudes to find {an} 
atom in the excited $\ket{2}$ and ground $\ket{1}$
states. Note that the norm of the wavefunction,
$|a|^2+|b|^2$, is not conserved {due to non-Hermitian 
term in the Hamiltonian.}

To study the atomic dynamics in the optical
lattice we use the standard Monte Carlo
wavefunction technique 
to get the most probabilistic outcome that can
be compared directly with corresponding experimental
observations with single atoms. 
The method is based on the evolution of {an} atomic state 
$\ket{\Psi(t)}$ while a continuous measurement of radiation-field state is performed 
by an ideal photodetector. The evolution consists of two parts: 
(1) jumps to the ground state ($a=0$, $|b|^2=1$) each of which is accompanied by the 
emission of an observable photon at random time moments 
with the mean time $(|a|^2\ \Gamma)^{-1}$ and (2) coherent 
evolution with continuously decaying norm of the atomic state vector
without the emission of an observable photon. 
The decay of the norm of the state vector is equal to the probability 
of spontaneous emission of the next photon.
This coherent decay without emission of a photon is 
usually interpreted within the context of {the} quantum measurement theory.
Let us introduce
the new real-valued variables {(normalized all the time)} instead of the 
amplitudes $a$ and $b$ {(renormalized only after SE events)}
\begin{equation}
\begin{displaystyle}
u\equiv\frac{2\Rre\left(ab^*\right)}{\left|a\right|^2+\left|b\right|^2},
\quad
v\equiv\frac{-2\Iim\left(ab^*\right)}{\left|a\right|^2+\left|b\right|^2},
\quad
z\equiv\frac{\left|a\right|^2-\left|b\right|^2}{\left|a\right|^2+\left|b\right|^2},
\end{displaystyle}
\label{uvz_def}
\end{equation}
which have the meaning of synphase and quadrature
components of the atomic electric dipole moment
and the population inversion, respectively. We
stress that the length of the Bloch vector, $u^2+v^2+z^2=1$,
is conserved.

The average atomic momentum is supposed to be large as compared
with the photon momentum $\hbar k_f$, so the translational
atomic motion can be treated classically using the
Hamilton equations. The whole atomic dynamics
is governed by the following Hamilton-Schr\"odinger
equations \cite{Acta}: 
\begin{equation}
\begin{aligned}
\dot x&=\omega_r p,\quad
\dot p=-u\sin x+\sum\limits_{j=1}^{\infty}p_j\delta(\tau-\tau_j),
\\
\dot u&=\Delta v+\frac{\gamma}{2}uz-u\sum\limits_{j=1}^{\infty}\delta(\tau-\tau_j),
\\
\dot v&=-\Delta u+2 z\cos x+\frac{\gamma}{2}vz-v\sum\limits_{j=1}^{\infty}\delta(\tau-\tau_j),
\\                                         
\dot z&=-2 v\cos x-\frac{\gamma}{2}(u^2+v^2)-(z+1)\sum\limits_{j=1}^{\infty}\delta(\tau-\tau_j),
\end{aligned}
\label{mainsys}
\end{equation}
where $x\equiv k_f\aver{\hat X}$ and $p\equiv \aver{\hat P}/\hbar k_f$ are the
normalized atomic centre-of-mass position and momentum, respectively.
The dot denotes differentiation with respect to
the normalized time $\tau\equiv \Omega t$. The {values of the} 
normalized
decay rate $\gamma\equiv \Gamma/\Omega$ and the recoil frequency
$\omega_r\equiv\hbar k_f^2/m_a\Omega\ll 1$ are fixed in this paper,
and the normalized detuning $\Delta\equiv(\omega_f-\omega_a)/\Omega$
is a variable parameter. In the
equations of motion (\ref{mainsys}) $\tau_j$ are random
time moments of SE events
and $p_j$ are random recoil momenta with the
values between $\pm 1$ in a one-dimensional case.
In terms of the normalized time $\tau$ the mean
frequency of SE events is equal to
$\gamma (z+1)/2$. At the time moments $\tau=\tau_j$, the
atomic variables change as follows: $p\to p+ p_j$,
$u\to 0,\ v\to 0,\ z\to -1$. The well-known effects of acceleration, deceleration
and velocity grouping have been successfully simulated with Eqs.
(\ref{mainsys}) in Ref. \cite{Acta}. In the present numerical simulation we take a cesium atom
with the working transition $6S_{1/2}-6P_{3/2}$ ($\lambda_a=852.1$ nm
and $\Gamma=3.2\cdot10^7$ Hz) interacting with a rather strong
field with the resonant Rabi frequency
$\Omega=10^{10}$ Hz. Thus, the corresponding normalized recoil
frequency is $\omega_r=10^{-5}$ and the spontaneous decay
rate is $\gamma=3.3\cdot10^{-3}$.

It follows from Eqs. (\ref{mainsys}) that the centre-of-mass
motion is described by the equation for a nonlinear
physical pendulum with a frequency modulation
caused by  coherent internal atomic dynamics
and random kicks of the momentum. Besides {SE recoils} any 
atomic trajectory is determined by the coherent evolution of the
synphase component of the electric dipole moment $u$
between the events of SE
and its jumps at random time moments $\tau_j$. These jumps are not small. 
In our recent paper
\cite{AP07} we developed a theory of atomic transport
in an optical lattice (in the absence of
SE) based on a specific behavior of the variable
$u$ which performs shallow and fast oscillations between
the nodes of the standing laser wave and changes
suddenly its value when atoms cross the nodes. The
theory predicts deterministic chaotic transport at
small values of the detuning $|\Delta|\ll 1$ whose
statistical properties are very well described by
a stochastic map {for
the deterministic variable $u$} \cite{AP07}.
In fact, atom moves in a rigid optical lattice
just like as in a random optical potential with
a complicated alternation of oscillations in
potential wells and flights over many
wells when it can change its direction of motion
many times. SE causes further
complication of this motion.

\section{Chaotic walking at small detunings, $|\Delta|\ll 1$}

\begin{figure}
\onefigure[width=0.48\textwidth,clip]{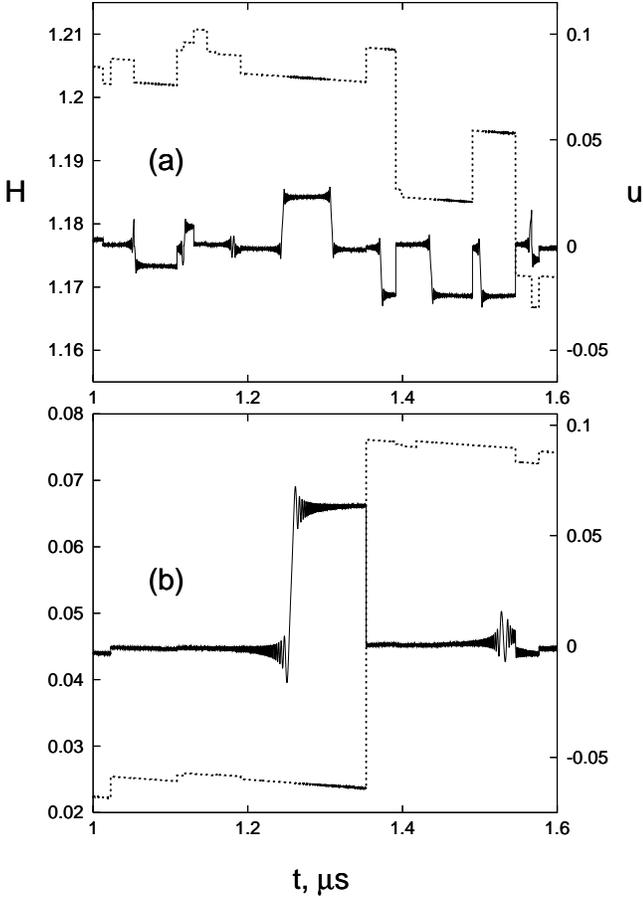}
\caption{Time evolution of the atomic energy $H$ {(dashed line)} and the synphase component of 
the electric dipole moment $u$0 {(solid line)} at the same small value of the 
detuning $\Delta=-0.001$ but 
with different initial conditions.}
\label{fig1}
\end{figure}

Near the resonance ($|\Delta|\ll 1$ ), the following quantity is
{almost} conserved between any two acts of SE:
\begin{equation}
\begin{aligned}
\tilde H_j &\equiv\frac{\omega_r}{2}p^2-u\cos x-\frac{\Delta}{2}z-\frac{\Delta\gamma}{4}\aver{1-z^2}(\tau-\tau_j)=\\
&=H-\frac{\Delta\gamma}{4}\aver{1-z^2}(\tau-\tau_j),
\end{aligned}
\label{H}
\end{equation}
where $\tau_j\leqslant\tau<\tau_{j+1}$.
$H$ is the total atomic energy which is a constant in the purely Hamiltonian system, i.e. 
 without any relaxation  \cite{AP07}.
The last term in (\ref{H}) with the averaging over a time exceeding the period
of the Rabi oscillations compensates the
relaxation. The energy $H$ changes suddenly at the
moments of SE and decays linearly
in between ({Fig.~\ref{fig1})} whereas the pseudoenergy $\tilde H$
changes suddenly as well at the same moments but is approximately
a constant in between. At $H\gtrsim |u|$, the atom moves
ballistically, and its momentum cannot be zero
because at small detunings $|\Delta|$ the
kinetic energy is larger than all the other
terms in (\ref{H}). If $ H\lesssim 0$, the atom changes
surely the sign of the momentum during its
motion. Thus, if we {could construct a} mapping
for the pseudoenergy $\tilde H_j$,
we {would} know approximately when
atoms move ballistically and when they turn
and could estimate the duration of the atomic flights
which is a time {interval} between two successive events
when atom changes the sign of $p$.

Just after SE at $\tau=\tau_j$, we 
have $u\to0$, $z\to-1$, $p\to p+p_j$ and $H\to\tilde H_j$,
and a change in the energy $H$ during the time {interval} between the acts of SE
is equal to the difference between the values of the pseudoenergy
just before ($\tilde H_{j-1}$) and just after the $j$-th 
SE ($\tilde H_j$) 
\be
\begin{aligned}
&H_j-H_{j-1}=\tilde H_j-\tilde H_{j-1}=\omega_r p(\tau_j) p_j+\frac{\omega_r}{2} p_j^2+\frac{\Delta}{2}+\\
&+u(\tau_j)\cos x(\tau_j)+\frac{\Delta}{2}z(\tau_j)+\frac{\Delta\gamma}{4}\aver{1-z^2}(\tau_j-\tau_{j-1}),
\label{DH}
\end{aligned}
\end{equation}
where $H_j$ is a value of the energy just after $\tau=\tau_j$ and 
$x(\tau_j)$, $u(\tau_j)$, $z(\tau_j)$, $p(\tau_j)$ are the values  
of the corresponding variables just before the moments $\tau=\tau_j$ 
which are determined by coherent evolution between SE events.
Changes in $H$ at $\tau=\tau_j $ are conditioned mainly by the
corresponding changes in $u$ (Fig.~\ref{fig1}). 
We stress that {sudden changes in $u$ occur at the moments of} 
crossing the nodes of the standing wave
and SE events. In Fig.~\ref{fig1}{b} the jump just after $t=1.2$ $\mu$s
occurs when the atom crosses a node whereas the jump just before
$t=1.4$ $\mu$s is caused by a SE event.

Let us estimate the average value of the energy jump's magnitude.
Analytical estimate and simulation show that with sufficiently large values
of the momentum, $\omega_r |p|\gtrsim\gamma/2$, and at small detunings, 
we get $\aver{u\cos x}\simeq\aver{z}\simeq 0$ during the coherent
evolution. The component $u$ never goes far away from zero, and 
$z$ performs frequency-modulated harmonic oscillations 
in the range $-1\lesssim z\lesssim 1$. The probability of SE is 
proportional to $(z+1)/2$. We estimate the average value of
the population inversion just before SE events at $\tau=\tau_j$ 
to be equal to $\aver{z(\tau_j)}\simeq 0.5$.
In the expression (\ref{DH}) the first and fourth
terms are estimated to be zero in average but the second,
third, fifth and sixth ones are not. The total average
change in $H_j$ due to SE and the relaxation term is
\be
\aver{H_j- H_{j-1}}=\frac{\omega_r}{6}+\Delta,
\end{equation}
where we estimated the average value of $z^2 $ to
be $1/2$ and the average value of the squared recoil momentum
$p_j^2$ is $1/3$. The mean time between SE
events is $\aver{\tau_j-\tau_{j-1}}=2/\gamma$.

Thus, the evolution of the energy $H$ is an
asymmetric random walking. At positive values of
the detuning $\Delta$, the values of $H$   
increase in average, whereas at $\Delta<0$ they may
increase or decrease depending on the relations
between $\omega_r$ and $\Delta$. We take the values of the
detuning to be $|\Delta|/2\gg\omega_r/9$, and if $\Delta<0$ then
$H_j$ decreases in average. The corresponding physical
effects --- light-induced acceleration and deceleration of atoms --- are well
known \cite{Kaz}. 

\begin{figure}
\onefigure[width=0.48\textwidth,clip]{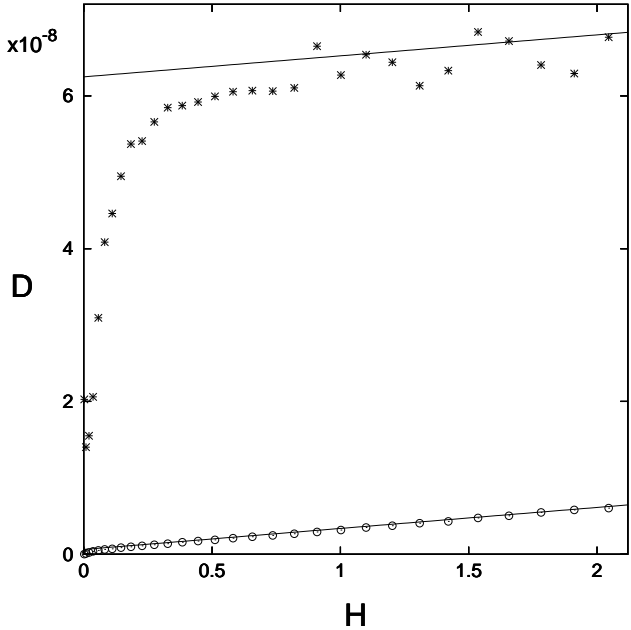}
\caption{{The diffusion coefficient $D$ in the energetic space
vs the energy $H$: stars --- $\Delta=-0.001$, circles --- $\Delta=-0.0001$
(numerical data). Solid lines show the corresponding analytical curves obtained
with the formula (\ref{D}).}}
\label{fig2}
\end{figure}

The diffusion coefficient in the
energetic space can be estimated with the help of the two largest
terms in Eq. (\ref{DH}) as
\be
\begin{aligned}
\ &D\equiv\frac{\aver{(H_j-H_{j-1})^2}-\aver{H_j-H_{j-1}}^2}{4\aver{\tau_j
-\tau_{j-1}}}\simeq\\
\ \simeq&\frac{\aver{\omega_r^2p^2p^2_j}\gamma+
\aver{u^2(\tau_j)\cos^2 x(\tau_j)}\gamma}{8}\simeq\frac{\omega_r H_j\gamma}
{12}+
\frac{\Delta^2}{{16}},
\label{D}
\end{aligned}
\end{equation}
where we suppose $u$ to be a random-like process ({Fig.~\ref{fig1}}) described
by the Eq. (11) from Ref.\cite{AP07}, which is correct only under the condition
$|p|\ll1/\omega_r$. This condition is fulfilled for all the atomic flights
found in the numerical simulation. By the other hand, there
is {a stronger} condition, $\omega_r|p|\gtrsim\gamma/2$, that
is needed to neglect correlations between $u^2$ and $\cos^2 x$. {Thus, 
the expression (\ref{D}) fails to give a result 
supporting the corresponding numerics for
sufficiently slow atoms. In Fig.~\ref{fig2} we compare numerical and 
analytical results 
(obtained with Eq. (\ref{D}) for the diffusion coefficient $D$ with two 
values of the detuning $\Delta=-0.001$ and $\Delta=-0.0001$.  
As expected, the correspondence is rather good for moderate
values of the energy but not for small values (slow atoms) when
the analytical formula (\ref{D}) is not applicable.}
The drift velocity of particles {in the energetic space} 
at sufficiently large values of $p$ is
\be
c\equiv\frac{d\aver{H}}{d\tau}=\frac{\aver{H_j-H_{j-1}}}{\aver{\tau_j-\tau_{j-1}}}
=\frac{\omega_r\gamma}{12}+\frac{\Delta\gamma}{2}.
\label{c}
\end{equation}
This result is also correct only with sufficiently fast atoms.
At the same condition {as (\ref{c})} we can estimate the friction force acting on the atom
to be the following:
\be
F\equiv\frac{d\aver{p}}{d\tau}\simeq\frac{\Delta\gamma}{2\omega_r p}.
\end{equation}
This kind of decreasing $F$ with increasing $p$ is a
well-known fact \cite{Kaz}.

Random jumps of the atomic energy just after
SE give rise to a random walking of atoms in an
optical lattice. In order to find distribution of
the durations of atomic flights we consider the problem
of the first passage time for the quantity $H_j$ to
return to its zero value (to be more correct, a 
return of $H$ to $|u|$ must be considered, but {with $|\Delta|\ll 1$ the 
variable} 
$u$ cannot reach large values and always returns to zero value
{for the}  time $\sim 2/\gamma$, see {Fig.~\ref{fig1}}). In the very beginning of any flight
we have $H_j\approx 0$, then it can reach a rather large value
and after that it returns to zero. The duration of this
process is a flight duration $T$.

If the random jumps of $H_j$ would be symmetric, the
probability to find the flight duration to be equal to $T$
would be proportional to $T^{-1.5}$, where the exponent
does not depend on the diffusion coefficient (a classical result in theory of 
symmetric random walking \cite{Feller}).

\begin{figure}
\onefigure{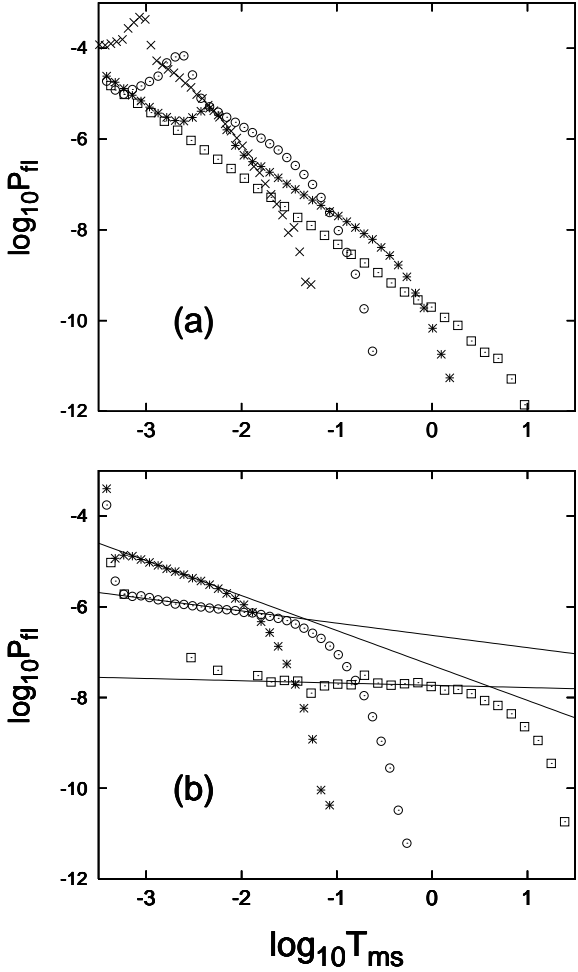}
\caption{The PDFs $P_{\rm fl}$ for the duration of atomic flights $T_{ms}
$
in milliseconds with 
(a) small detunings (crosses $\Delta=-0.01$, stars $\Delta=-0.001$, 
circles $\Delta=-0.0001$, squares $\Delta=-0.00001$) and 
(b) medium detunings (stars $\Delta=-0.09$, $\alpha = -0.77$; 
circles $\Delta=-0.1$, $\alpha = -0.27$; squares $\Delta=-0.12$, 
$\alpha = -0.05$). Straight lines show slopes $\alpha$
of the power-law fragments of the PDFs in log-log scale. }
\label{fig3}
\end{figure}

What will happen if we take into account that
the random walking of $H_j$ is asymmetric? At $\Delta>0$,
atoms  begin  to  accelerate  without any
flights. At $\Delta<0$, the friction force tends to stop
atoms, and instead of the power-law decay $T^{-1.5}$ we get an
exponential one at large $T$. At large times exceeding
the mean SE time $2/\gamma$, one may treat the evolution of $H$
as a diffusion process with a drift described
by the Fokker-Planck equation in the energetic
space
\begin{equation}
\dot P(H, \tau)=-2c\frac{\partial P}{\partial H}+ D\frac{\partial^2 P}{\partial H^2},
\end{equation}
where the diffusion $D$ and drift $c$ coefficients are given 
by Eqs. (\ref{D}) and (\ref{c}), respectively.
If the coefficients $c$ and $D$ would not change with changing the energy, the
PDF for flight durations would be equal to \cite{Feller}
\begin{equation}
P_{\rm{fl}}\propto e^{-c^2T/ D}T^{-1.5}.
\label{stat}
\end{equation}
This result agrees qualitatively with the results of
numerical simulation shown in Fig. {\ref{fig3}}a
for a few values of the detuning $\Delta$.
Really, all the PDFs in Fig.~{\ref{fig3}}a have power-law fragments followed by exponential
tails at large $T$ in accordance with the formula (\ref{stat}). However,
the length of these fragments depends strongly on the value of the detuning.
At very small value $\Delta=-10^{-5}$ (when coherent atomic dynamics is
practically regular \cite{AP07} and atom performs a random walk due to SE),
$P_{\rm{fl}}\sim T^{-1.5}$, whereas at larger values of $|\Delta|$ the power-law
fragments are much shorter.
In fact, both $c$ and $D$ depend on the value of $H$, {therefore,} 
Eqs. (\ref{D}) and (\ref{c}) are not correct for small values of
the momentum $p$, and more accurate formula for
$P_{\rm{fl}}$ is {required}.

\begin{figure}
\onefigure[width=0.48\textwidth,clip]{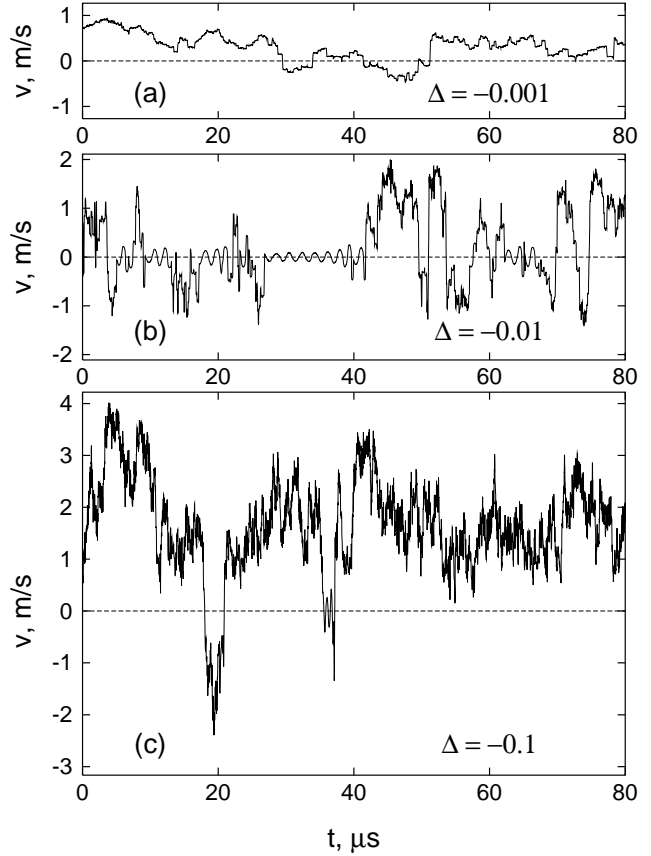}
\caption{Dependencies of the current atomic velocity $v$ on time: 
(a) $\Delta=-0.001$, chaotic walking with comparatively long flights, 
(b) $\Delta=-0.01$, chaotic walking with short flights, 
(c) $\Delta=-0.1$, velocity grouping effect.}
\label{fig4}
\end{figure}

To illustrate the behavior of the atomic momentum
at different values of the detuning we plot in Fig.~{\ref{fig4}}a a typical
chaotic walking of an atom with
comparatively long flights at $\Delta=-0.001$ and
in Fig.~{\ref{fig4}}b chaotic walking with short
flights at $\Delta=-0.01$. 

Qualitatively, the statistics in Fig.~{\ref{fig3}} are similar to those 
computed in Ref. \cite{AP07} with a purely Hamiltonian coherent dynamics.  
{Both of} the PDFs contain power-law fragments with exponential 
decays at their tails, but the origin of these fragments and tails is different. 
Statistics {of the} purely Hamiltonian system \cite{AP07} is 
{governed} by   
a deterministic diffusion of the quantity $u$ in a bounded space (between its 
maximal and minimal values $\pm 1$), {\it at a constant energy $H$}. 
Solution for the first-passage-time problem for the quantity  $u=\pm H$ 
gives a PDF with a power-law fragment with the slope $-1.5$ and an 
exponential tail. The quantity $u$  
jumps to zero value at the moments of SE (see {Fig.~\ref{fig1}}), 
and its evolution cannot be treated as a diffusion process when we take into 
account SE. However, the energy now is not a constant and can walk randomly 
(with a drift) within a broad range. The statistics of the system with SE 
is defined by   
a random walking of the energy $H$, not $u$. Thus, in both the systems the condition 
$H\approx\pm u$ defines the value of the energy which allows atoms to stop and turn back. 
Transformation of power-law fragments into exponential tails is explained 
in the purely Hamiltonian system \cite{AP07} by {a limitation} of the 
quantity $u$, 
whereas in the system with SE it is explained by a drift of the energy $H$. 
{Both of  those} factors prevent randomly walking quantities to go far from their 
critical values and decrease exponentially the probability of long atomic 
flights.

\section{Chaotic walking at moderate detunings, $|\Delta|\lesssim 1$}

The effect of velocity grouping, when there are one or a few values
of the capture momentum $p_g$ to which current momenta 
of different atoms tend to, is known to occur
at moderate negative values of the detuning \cite{Kaz}.
This effect has been numerically demonstrated
with Eqs. (\ref{mainsys}) in our recent paper \cite{Acta}.
Chaotic walking of atoms may occur in the regime
of velocity groping if the values of $p_g$ are sufficiently small {and  
atoms} can change the direction of motion {due to} fluctuations
of momentum (caused by chaos in coherent evolution and/or SE, see Fig. {\ref{fig4}}c).

Computed statistics of atomic flights at medium values of the detuning $|\Delta|\lesssim 1$
(Fig.~{\ref{fig3}}b) {are} similar to that at small detunings
but the slope of the power-law decay may differ
considerably. The PDFs $P_{\rm{fl}}(T)$ shown in Fig.~{\ref{fig3}}b
demonstrate decrease of the slope with increasing
the values of $|\Delta|$ with corresponding increase of the
lengths of the {power-law} fragments. It should be emphasized
that the length of the power-law fragments (and the mean
flight length) increases significantly with a rather small {increase}
in $|\Delta|$.

\begin{figure}
\onefigure[width=0.48\textwidth,clip]{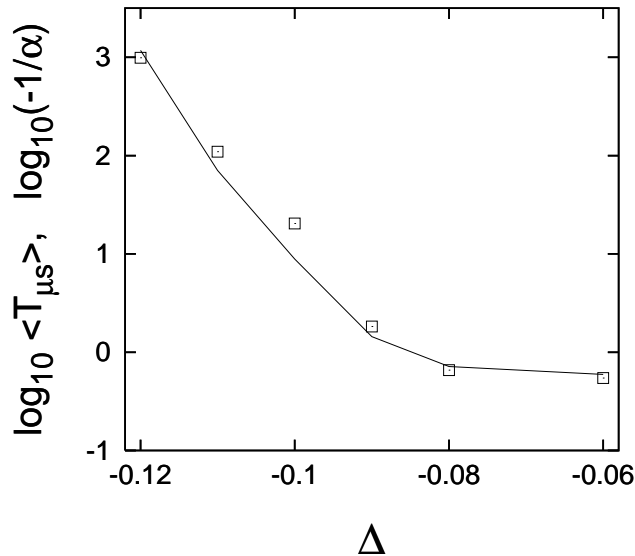}
\caption{Dependencies of the logarithms of the mean duration of
atomic flights $\langle T_{{\mu} s}\rangle$ (in microseconds) (solid line)
and of the slope $\alpha$ of the PDF power-law fragments (squares)
on the medium detuning $\Delta$.}
\label{fig5}
\end{figure}

Varying the value of the detuning $\Delta$, one can manipulate 
atomic transport in an optical lattice and its statistical properties. 
In Fig. {\ref{fig5}} we plot the dependencies of the mean duration of 
atomic flights $\langle T\rangle$ (in microseconds) and the slope of the power-law fragments of 
$P_{\rm fl} \sim T^{\alpha}$ on  the detuning in the range of medium values. 
The dependencies correlate well with each other. Figure {\ref{fig3}~}b 
demonstrates clearly that the length of the power-law fragments increases with increasing 
the absolute values of the detuning, whereas the absolute value of the slope 
$\alpha$ decreases. The mean duration of flights $\langle T\rangle$ 
increases correspondingly with increasing $|\Delta|$.

The control is nonlinear 
in the sense that when slightly decreasing $\Delta$, say, from $-0.08$ 
to $-0.12 $, the mean time of flights increases in a few orders of 
magnitude (see Fig.~ {\ref{fig5}}). This effect is explained by 
{increase in} the 
value of the capture momentum $p_g$ and decreasing fluctuations of the 
current atomic momentum {with increase in the absolute value of 
$|\Delta|$.} For example, at $\Delta=-0.1$, 
one gets $p_g \simeq 500$ ($\simeq 1.5$ m/s) with the momentum fluctuations 
of the same order. Thus, the atom can change its direction of motion (see 
Fig. {\ref{fig4}~}c,
with $\Delta=-0.1$ the mean time of flights {is} $T_{\mu s}\sim 10 \mu$s) but not 
so frequently as in the case of smaller detunings (see Fig. {\ref{fig4}b}). 
The value of $p_g$  increases with increasing $|\Delta|$ and, say, 
at $p_g \simeq 1000$ 
the momentum fluctuations are of the order of $300$, and the atom cannot 
change its direction of motion. {Reduction in the } momentum fluctuations is caused by 
{increase in} the friction force $F$. With the values of $p$ smaller than $p_g$,  
the force $F$ is so large that changes in the energy $H$ with time is not 
{a}  
process of random walking (as it is in the case with smaller detunings) 
but rather a directed drift in the momentum space to the value of the 
capture momentum. Thus, it is practically impossible for atoms to decrease their 
current values of $p$ to zero value, and the process of chaotic walking 
eventually stops. The slope $\alpha$ of the PDF
power-law fragments can go to zero due to existence of an exponential 
decay at the very tail of $P_{\rm fl}$.  With a purely power-law decay 
the minimal slope would be $\alpha=-1$.

In conclusion, we have shown that near the resonance atomic transport in 
an optical lattice is a complicated process of chaotic walking caused by 
an interplay between coherent but deterministically chaotic internal
atomic dynamics and spontaneous emission 
random events. It is possible to manipulate this process and its 
statistical characteristics by varying the single control parameter,  
the atom-laser detuning $\Delta$.  

This work was supported  by the Russian Foundation for Basic Research
(project no. 06-02-16421) and by the Presidential grant no. MK -- 1680.2007.2.

\end{document}